\documentclass[superscriptaddress,
 amsmath,amssymb,
 aps,twocolumn,
apl,
]{revtex4-2}

\usepackage{graphicx}
\usepackage{dcolumn}
\usepackage{bm}
\usepackage{color}
\usepackage{hyperref}
\usepackage{multirow}

\begin{document}


\title{Padé Approximation and Partition Function Zeros}

\author{R. G. M. Rodrigues}
\email{rgmr@unicamp.br}
\affiliation{Instituto de Computação - Universidade Estadual de Campinas (UNICAMP), 13083-852, Campinas, São Paulo, Brasil}

\begin{abstract}


Fisher zeros play a central role in the theoretical understanding of phase transitions. However, their computation requires knowledge of the density of states, which limits their practical applicability. Alternative approaches based on the Energy Probability Distribution (EPD) and Moment Generating Function (MGF) alleviate the computational cost but suffer from convergence issues in the two-dimensional \textbf{anisotropic Heisenberg} model (XY model). In this work, we introduce a Padé approximation to systematically reduce the number of zeros required in the Fisher, EPD, and MGF formulations without loss of accuracy. Moreover, since the Fisher zeros formulation does not rely on a convergence algorithm, combining this approach with a Padé approximation enables a reliable analysis of the XY model while significantly reducing computational cost. Applications to the two-dimensional Ising and XY models demonstrate substantial decreases in polynomial degree and computation time while preserving accurate estimates of the critical temperature.

\end{abstract}


\maketitle


\section{Introduction}

Partition function zeros provide a fundamental framework for studying phase transitions in statistical physics. Following the seminal works of Yang and Lee \cite{yanglee1, yanglee2}, and later Fisher \cite{fisherzeros}, the analytic continuation of thermodynamic parameters, such as magnetic field or temperature, into the complex plane allows the partition function to be expressed as a polynomial, whose zeros encode critical behavior. A phase transition is characterized by the accumulation of zeros near the real axis at the critical point. For finite systems, the zeros closest to the real axis, referred to as dominant zeros, determine the leading critical properties and move systematically toward the real axis as the system reaches the thermodynamic limit. This formalism has been extensively applied to a wide range of physical systems and has more recently been realized in experimental settings~\cite{rochapol, dynamicquant, benareview, graphzerosising, fisherzerosbosegas, Moueddene_2024, Moueddene_2025, tensorxylj, qcdcumulutant, expzerosyl, zerosmeasurequantsimul}.

While the Fisher zeros provide a rigorous foundation for the study of phase transitions, its computational implementation faces significant challenges. The method requires the solution of a high degree polynomial with coefficients given by the system density of states (DOS), a quantity often difficult to determine. Even when reliable estimates of the DOS are available, its values typically span many orders of magnitude, leading to severe numerical instabilities, including underflow, overflow, and loss of precision due to round-off errors. These problems can be addressed using alternative formulations such as the Energy Probability Distribution (EPD) zeros \cite{epd}, the Moment Generating Function (MGF) zeros \cite{mgf}, and the cumulant method \cite{cumulants}. These methods exploit information contained in the energy probability distribution (EPD), energy moments (MGF), or energy cumulants (cumulant method) to infer the critical temperature. Since each method relies on partial statistical information about the system, an iterative refinement algorithm is employed to ensure convergence toward the true critical temperature. These approaches have demonstrated high accuracy and robustness in the analysis of discontinuous, continuous, and topological phase transitions \cite{epd, mgf, Lima2019, Costa2019, cumulants, cumulants1, cumulants2}.

Compared to the Fisher zeros approach, the EPD and MGF formalisms involve polynomials of significantly lower degree and yield a reduced set of zeros in the complex plane, thereby alleviating many numerical instabilities associated with the direct computation of Fisher zeros. However, when applied to the two-dimensional XY model, they face fundamental challenges: their convergence algorithms, designed to iteratively refine an initial temperature guess toward the critical point, become unreliable and fail to converge consistently, thus preventing the identification of the phase transition. In this work, we address this issue by introducing a Padé approximation to the Fisher zeros method, enabling an accurate analysis while retaining a reduced set of zeros. Furthermore, this approach extends naturally to the EPD and MGF formalisms, further reducing the number of zeros without loss of accuracy and with significantly reduced runtime.

This paper is organized as follows. Sections~\ref{sec:fisher}, \ref{sec:epd}, and \ref{sec:mgf} provide brief reviews of the Fisher, EPD, and MGF zeros methods, respectively, while the Padé approximation is introduced in Section~\ref{sec:pade}. The Ising and XY models are described in Sections \ref{sec:ising} and \ref{sec:xy}. Sections~\ref{res:pade_fisher}, \ref{res:pade_epd}, and \ref{res:pade_mgf} then present the application of the Padé approximation to the Fisher, EPD, and MGF zeros in the Ising model. Section~\ref{res:problemxy} discusses the limitations of the EPD and MGF approaches in the analysis of the XY model, and Section~\ref{res:models} reports the final results for both the XY and Ising models.

\section{\label{sec:fisher}Fisher Zeros}

The concept of Fisher zeros ~\cite{fisherzeros} constitutes an extension of the Yang–Lee theory to the canonical ensemble. Fisher’s formulation extends the inverse temperature, $\beta = 1/(k_{\mathrm{B}}T)$, to complex values. In this framework, the canonical partition function can be expressed as
\begin{equation}
    Z = \sum_E g(E)e^{-\beta E}
      = e^{-\beta \varepsilon_0}\sum_{k} g_{k} z^{k},
    \label{eq:Fisher}
\end{equation}
where $z = e^{-\beta \varepsilon}$, $k_{\mathrm{B}}$ is the Boltzmann constant, and $g(E)$ denotes the density of states. The energy spectrum is discretized as $E_k = \varepsilon_0 + k\varepsilon$, with $\varepsilon_0$ the ground state energy, $\varepsilon$ the energy bin, and $g_k = g(E_k)$ the corresponding density of states value~\cite{rochabkt}. This polynomial has complex conjugate roots, and since $g(E) \geq 0$, any real root must be negative. A real positive root appears only in the thermodynamic limit, as the dominant zero approaches the positive real axis, signaling the point where the free energy develops a non-analyticity and a phase transition occurs.

\section{\label{sec:epd}The Energy Probability Distribution Zeros}

The Energy Probability Distribution zeros method~\cite{epd} provides an alternative to the Fisher zeros, obtained by rescaling the variable $z$ and expressing the partition function in terms of the energy probability distribution. This reformulation allows the zeros to be determined from a histogram rather than from the density of states, thus simplifying the numerical implementation while preserving the underlying analytical structure.

The EPD polynomial is found by multiplying Eq.~\ref{eq:Fisher} by the identity $e^{-\beta_o E}e^{\beta_o E} = 1$, which gives
\begin{equation}\label{eq:epd}
    Z = \sum_E g(E)e^{-\beta_o E}e^{-\Delta \beta E} = e^{-\Delta \beta \varepsilon_o} \sum_k h_{\beta_o}(k) x^k,
\end{equation}
where $\beta_o = 1/(k_{\mathrm{B}}T_0)$ is an arbitrary reference inverse temperature, $\Delta\beta = \beta - \beta_o$, and $x = e^{-\Delta\beta \varepsilon}$. The coefficients $h_{\beta_o}(k) = g_k e^{-\beta_o E_k}$ correspond to the unnormalized energy probability distribution evaluated at $\beta_o$.  The polynomial degree is reduced by introducing a cutoff threshold $h_t$ that removes coefficients associated with negligible probabilities, while preserving the zeros relevant to the thermodynamics behavior at $\beta_o$.

Because the EPD zeros are a rescaled version of the Fisher zeros through the transformation $x = z / e^{-\beta_o \varepsilon}$, the Fisher dominant zero $z_c = e^{-\beta_c \varepsilon}$ maps onto $x = 1$ when $\beta_o = \beta_c$. This property enables an iterative algorithm that refines an arbitrary initial $\beta_o(L)$ toward $\beta_c(L)$. The algorithm is given by

\begin{enumerate}
    \item Construct a single histogram $h_{\beta_{0}}^{j}$ at $\beta_{0}^{j}$ and rescale it so that max$(h_{\beta_{0}}^{j}(k)) = 1$.
    \item Discard the coefficients $h_{\beta_{0}}^{j}(k)$ that are smaller than a chosen threshold $h_t$.
    \item Find the zeros of the polynomial with coefficients given by $h_{\beta_{0}}^{j}$.
    \item Find the dominant zero, $x_{c}^{j}$.
    \begin{enumerate}
        \item[(a)] If $x_{c}^{j}$ is close enough to the point $(1,0)$, stop.
        \item[(b)] Else, make 
        \[
        \beta_{0}^{j+1} = -\frac{\ln(\Re e\{x_{c}^{j}\})}{\varepsilon} + \beta_{0}^{j}
        \]
        and go back to 1.
    \end{enumerate}
\end{enumerate}

\section{\label{sec:mgf} Moment Generating Function Zeros}

The moment-generating function zeros \cite{mgf} are obtained by expanding the exponential term in the EPD polynomial, Eq.~\ref{eq:epd}, as a power series and exchanging the order of summation. Through this expansion, the polynomial coefficients become the moments of the energy probability distribution evaluated at $\beta_o$. The canonical partition function can thus be expressed as
\begin{equation}\label{mgfz}
    Z(\beta) = Z(\beta_o)\sum_{k=0}^{\infty} \frac{O_k(\beta_o)}{k!}y^k,
\end{equation}
where $Z(\beta_o)$ is a constant, $y=-\Delta\beta$, and $O_k(\beta_o) = \langle E^k \rangle_{\beta_o}$ are the moments of the energy. The MGF zeros encode the same information as the Fisher zeros since $Z(\beta_o)$ is a constant. Its degree is reduced by truncating the series at a finite order $k_{\text{max}}$.

As in the EPD formulation, the MGF method also employs a convergence algorithm. From the relation $x = e^{y \varepsilon}$, one observes that when $\beta_o = \beta_c$, the dominant zero in the EPD approach is located at $x_c = 1$. Consequently, in the MGF representation, the dominant zero must lie at $(0,0)$. This correspondence establishes the convergence criterion, from which the iterative algorithm is defined as

\begin{enumerate}
    \item[1] Find the energy moments $O_k(\beta_o^j) = \langle E^k \rangle$ at $\beta^j_o(L)$.
    \item[2] Find the zeros of the polynomial with coefficients given by $O_k(\beta_o^j) / k!$.
    \item[3] Find the dominant zero, $y^j_c(L)$.
    \begin{enumerate}
        \item[a)] If $y^j_c$ is close enough to the point $(0, 0)$, stop.
        \item[b)] Else, make $\beta^{j+1}_o(L) = \beta^j_o(L) - \Re e\{y^j_c(L)\}$ and go back to $1.$.
    \end{enumerate}
\end{enumerate}

\section{Padé Approximation}\label{sec:pade}

Inspired by the MGF approach, which estimates the zeros using a truncated Taylor expansion, we employ a Padé approximation to represent the partition function. This technique expresses a function as a ratio of two polynomials, yielding a rational approximation that typically outperforms truncated Taylor series, particularly in the presence of poles.

To construct the Padé approximation, we consider a function $f$ represented by a power series:
\begin{equation}
f(r) = \sum_{k=0}^{K} a_k r^k.
\label{eq:basci_exp}
\end{equation}
The Padé approximation of order $[m/n]$ is defined as
\begin{equation}
F_{m,n}(r) = \frac{P_m(r)}{Q_n(r)}= \frac{\sum_{i=0}^{m} p_i r^i}{1 + \sum_{j=1}^{n} q_j r^j},
\label{eq:pade}
\end{equation}
where the coefficients $\{p_i\}$ and $\{q_j\}$ are determined so that the Taylor expansion of $F_{m,n}(r)$ agrees with $f(r)$ up to the highest possible order, i.e.,
\begin{equation}
f(r) - F_{m,n}(r) = \mathcal{O}(r^{m+n+1}),
\end{equation}
where $(m+n+1) \leq K$. Detailed derivations of the procedure used to determine the coefficients $\{p_i\}$ and $\{q_j\}$ required to construct the Padé approximation can be found in ~\cite{numrec}. For the present discussion, it suffices to say that the coefficients $\{p_i\}$ and $\{q_j\}$ are fully determined once the set $\{a_k\}$ is known. 

The Padé approximation is applied to each of the previously introduced methods by identifying the coefficients $a_k$ with the corresponding physical quantities. The resulting polynomials $P_m(r)$ and $Q_n(r)$ have distinct interpretations. The zeros of $P_m(r)$ correspond to the zeros of interest such as the Fisher, EPD, or MGF zeros, while the zeros of $Q_n(r)$ signal the presence of spurious poles, marking regions where the Padé approximation loses reliability. In particular, the parameter $m$ sets the number of partition function zeros, and in the asymptotic regime of large $m$ and small $n$, the full set of zeros is recovered. This flexibility enables the selective reconstruction of a subset of zeros while preserving the overall accuracy of the approximation. 

The general Padé algorithm, applicable to all formulations, is summarized as follows:

\begin{enumerate}
    \item At iteration $j$, construct the coefficients $a_k^{(j)}$ according to the chosen method:
    \begin{itemize}
        \item Fisher: $a_k^{(j)} = g_k$,
        \item EPD: $a_k^{(j)} = h_{\beta_o^{j}}(k)$,
        \item MGF: $a_k^{(j)} = O_k(\beta_o^{j})/k!$.
    \end{itemize}

    \item Construct the Padé polynomials $P_m(r)$ and $Q_n(r)$ using $a_k^{(j)}$.

    \item Compute the roots of $P_m(r)$ and $Q_n(r)$.

    \item Identify and discard spurious zeros of $P_m(r)$ that lie in close proximity to the zeros of $Q_n(r)$.

    \item Use $P_m(r)$ to determine the dominant zero according to the chosen method:
    \begin{itemize}
        \item Fisher: the root closest to the positive real axis,
        \item EPD: the root closest to $(1,0)$,
        \item MGF: the root closest to $(0,0)$.
    \end{itemize}
    
    \item (EPD and MGF only) If the convergence criterion is satisfied, stop; otherwise, update the reference parameter and repeat from step 1.

\end{enumerate}

\subsection{Parameters Choice}

\begin{figure*}
    \centering
    \includegraphics[width=0.49\linewidth]{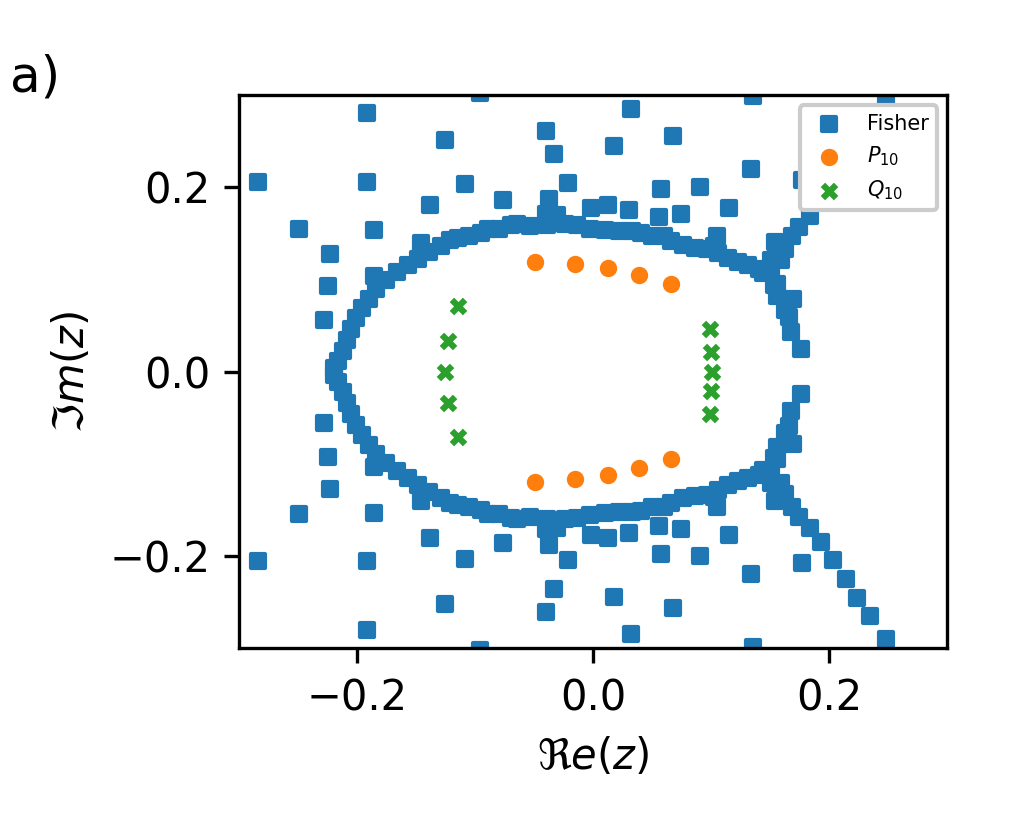}
    \includegraphics[width=0.49\linewidth]{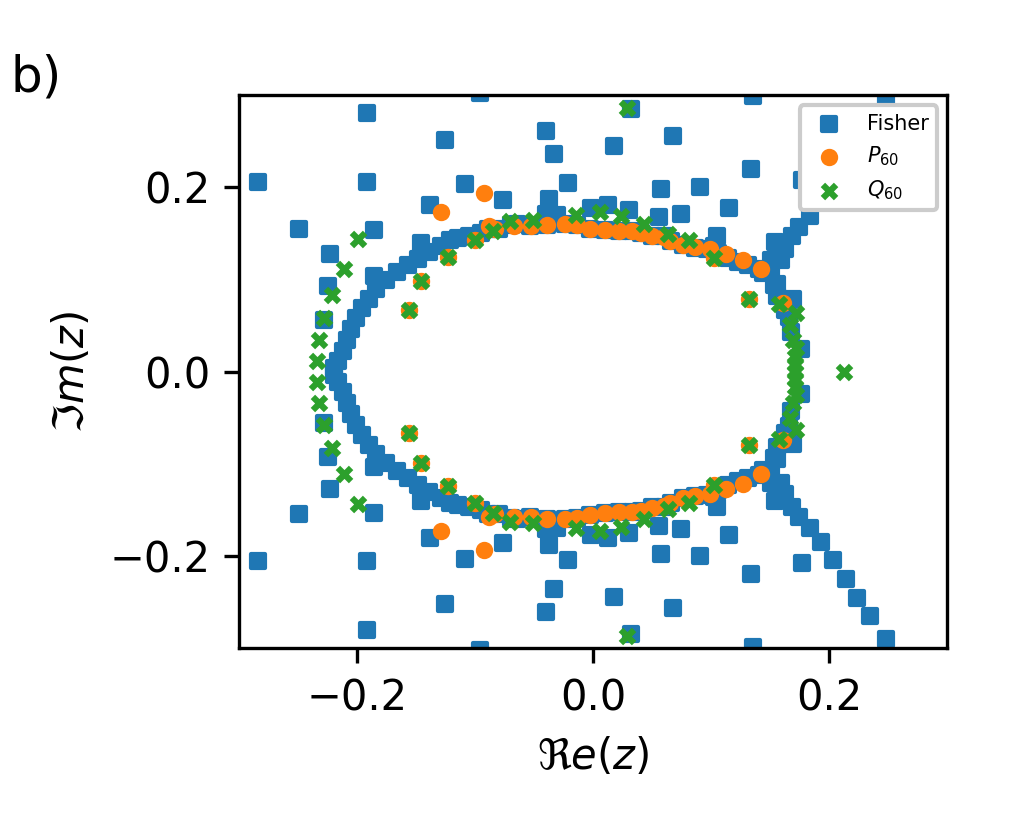}
    \includegraphics[width=0.49\linewidth]{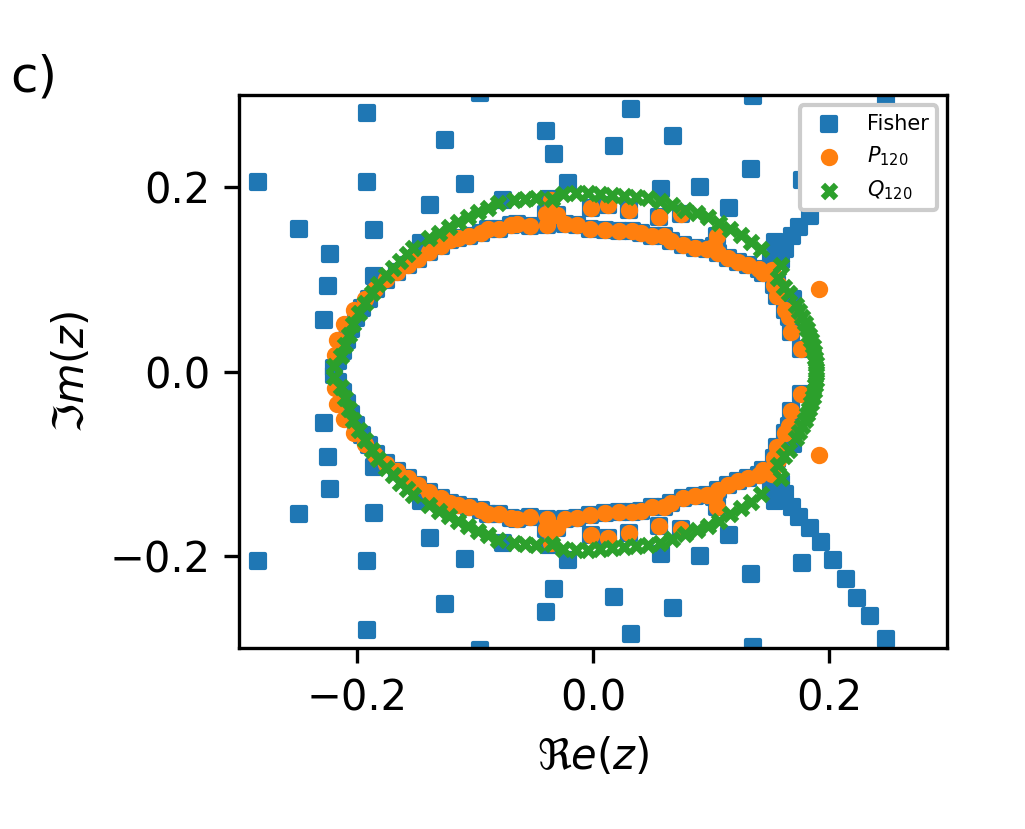}
    \includegraphics[width=0.49\linewidth]{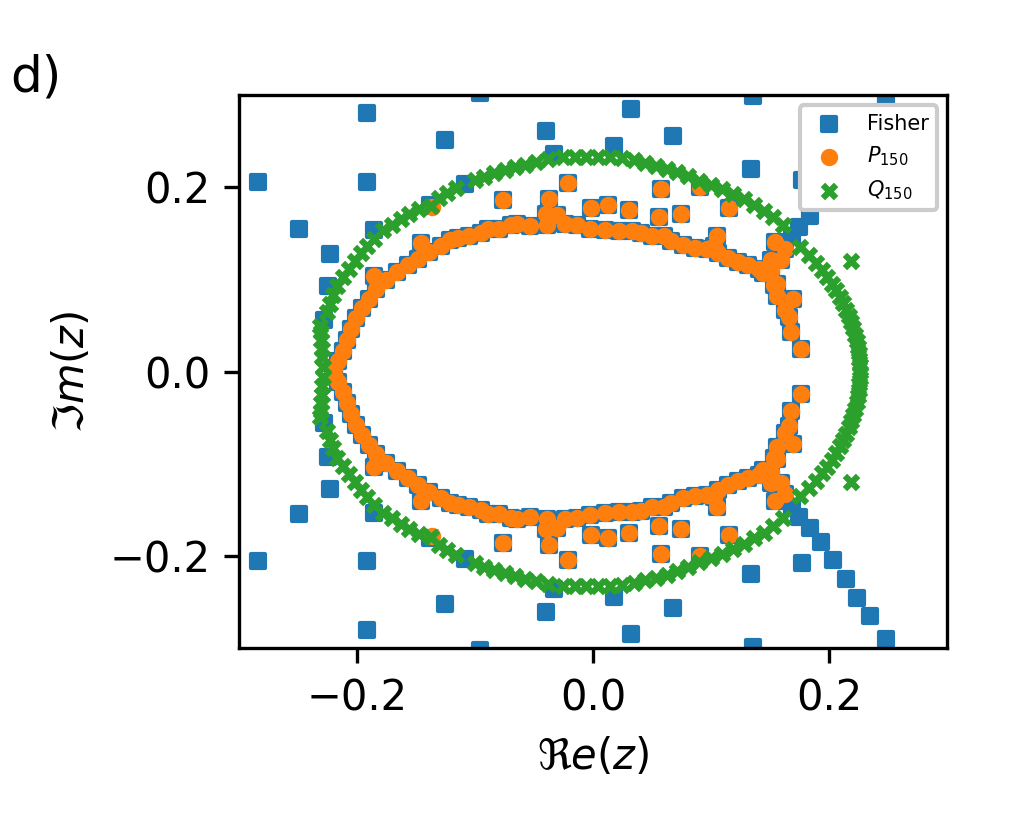}
    \caption{Fisher zeros for the Ising model with lattice size $L=24$, together with the Padé approximation for $m = n =[10,\,60,\,120,\,150]$. Figures (a,b) show that for small values of $m$ the approximation fails to accurately capture the region containing the dominant zero, whereas in (c,d) larger values yield an accurate reconstruction of the dominant zero and its nearby zeros. As $m$ increases, the instability region, indicated by the zeros of $Q_n(r)$, shifts away from the critical region.}
    \label{fig:pade_fisher}
\end{figure*}

To illustrate the role of the Padé parameters, we consider the Fisher zeros of the Ising model. As shown in Fig. \ref{fig:pade_fisher}, as $m$ increases, the Padé approximation progressively reconstructs the full set of Fisher zeros. At the same time, spurious poles from $Q_n(r)$ are displaced away from the zeros of $P_m(r)$.

In practice, the choice of $m$ cannot be determined a priori. A useful strategy is to calibrate $m$ using a smaller lattice size, selecting the minimum value that recovers the dominant zero, and then increasing it for larger systems.

\subsection{Shifted Padé Approximation}\label{sec:spade}

In practice, the quality of a series expansion can be significantly improved by centering the expansion around a point of interest, rather than the origin. In particular, when the relevant physical information is concentrated near a specific complex value $r_0$ (for instance, close to a dominant zero of the partition function), it is advantageous to first shift the expansion variable and then apply the Padé procedure.
 
Consider a generic truncated power series representation of a function $f$,
\begin{equation}
f(r)=\sum_{v=0}^{V} a_v r^v,
\label{eq:orig_series}
\end{equation}
where the coefficients $a_v$ are known from one of the formulations discussed previously (e.g., Fisher zeros, EPD, or MGF). We rewrite this polynomial as an expansion around a shifted point $r_0$ by introducing the variable $w = r - r_0$.

Using the binomial expansion, the series can be expressed as
\begin{equation}
f(r)=\sum_{k=0}^{V} c_k w^k,
\label{eq:shifted_series}
\end{equation}
with coefficients
\begin{equation}
c_k = \sum_{v=k}^{V} a_v \binom{v}{k} r_0^{\,v-k}, \qquad k=0,1,\dots,V.
\label{eq:shifted_coeffs}
\end{equation}
Once the shifted coefficients $\{c_k\}$ are obtained, we construct a Padé approximation of order $[m/n]$ in the variable $w$. The interpretation given before for the resulting polynomials $P_m$ and $Q_n$ remains unchanged. In terms of the algorithm, this modification affects only the coefficient construction step:

\begin{itemize}
    \item Construct the coefficients $a_k^{(j)}$ and compute the shifted coefficients $c_k^{(j)}$.
    \item Use $c_k^{(j)}$ to construct $P_m(w)$ and $Q_n(w)$.
\end{itemize}

Although this strategy can, in principle, be applied to the MGF and EPD zeros, its practical usefulness is limited by the difficulty of predicting the location of the dominant zero during the convergence process. A reliable estimate is only available at the transition temperature, where the dominant zero is expected to lie close to the origin for the MGF formulation and near the point $(1,0)$ for the EPD formulation. For this reason, in the following sections we restrict the application of the shifted Padé approximation to the Fisher zeros.

\subsection{Computational Challenges in Shifted Padé Algorithm}\label{sec:shifted_pade}

The shifted construction offers a significant advantage when applied to the Fisher zeros method. However, a naive implementation of the algorithm used to compute the coefficients $c_k$, as defined in Eq. \ref{eq:shifted_coeffs}, would make the method slower than previously discussed approaches.

To apply the Shifted Padé method to the Fisher zeros, it is necessary to calculate the coefficients given by
Eq. \ref{eq:shifted_coeffs} with $a_v = g_v$. This computation presents two main challenges. First, numerical instability may arise from overflow or underflow during the evaluation of the terms $g_v \binom{v}{k} r_0^{\,v-k}$, since the combinatorial factor and the density of states can attain extremely large values, while $r_0^{v-k}$ may become very small for values of $r_0$ between $0$ and $1$. Second, the direct evaluation of the summation is computationally expensive since $V$ can be large and there is a binomial coefficient that must be computed. Therefore, this shifts the computational cost from the root-finding algorithm to the coefficient construction itself.

To achieve both improved performance and a fair comparison between algorithms, the coefficients for the shifted Padé method were not computed using \texttt{Python}, as is done in the other methods. Instead, the algorithm was implemented in \texttt{Fortran}, using Horner’s method in combination with the arbitrary-precision \texttt{MPFUN} library~\cite{BaileyMPFUN}. This choice is motivated by the need for a consistent comparison with the dominant computational component of the other approaches, namely the root-finding procedure, which relies on the \texttt{MPSolve} library implemented in \texttt{C}. In this way, the runtime comparison is performed between implementations in \texttt{C} and \texttt{Fortran}, rather than against \texttt{Python}, thereby providing a more balanced and meaningful assessment.

\section{Ising Model}\label{sec:ising}

The two-dimensional Ising model was selected as a test system due to the availability of its exact density of states, which can be computed numerically \cite{exactising}. The model is on a square lattice, where each site 
$i$ has a spin variable $\sigma_i=\pm 1$, and the interaction between nearest-neighbors is governed by the Hamiltonian,
\begin{equation}
H = -J\sum_{\langle i,j \rangle} \sigma_i\sigma_j,
\end{equation}
where $J$ is a coupling constant and the summation runs over all pairs of nearest-neighbors. The model exhibits a second order phase transition at the exact critical temperature $T_c = 2 / \ln(1+\sqrt{2})J/k_B$. This model serves as a benchmark for validating the accuracy between the methods presented in this work. Throughout this study, the constants are fixed as $J=1$ and $k_\text{B} = 1$.

\section{Anisotropic Heisenberg or XY Model}\label{sec:xy}

The two-dimensional \textbf{anisotropic Heisenberg} model provides a paradigmatic example of a system undergoing a Berezinskii-Kosterlitz-Thouless (BKT) transition \cite{kt1, kt2}, which has been the subject of sustained theoretical and experimental interest for more than four decades~\cite{40yearskt}. Unlike conventional first- or second-order phase transitions, the BKT transition is topological in nature and does not involve spontaneous symmetry breaking or the emergence of true long-range order \cite{nolongrangeorderkt}. Instead, the low-temperature phase is characterized by quasi-long-range order, with two point correlation functions exhibiting an algebraic decay for $T \leq T_{\mathrm{BKT}}$, while above the transition temperature the correlations decay exponentially.

The transition occurs at a characteristic temperature $T_{\mathrm{BKT}}$, below which the system exhibits a line of critical points extending down to zero temperature. All derivatives of the free energy remain finite at the transition, and for this reason the BKT transition is classified as an infinite order phase transition. Despite the absence of divergences, the free energy is non-analytic for $T\leq T_{\mathrm{BKT}}$ and, in the thermodynamic limit, it is expected to see a line of zeros in the Fisher zeros map along the real temperature axis, while for $T > T_{\mathrm{BKT}}$ the zeros should not touch the real positive axis.

To establish notation and provide the basis for the numerical analysis that follows, the XY model is described by the Hamiltonian
\begin{equation}
\mathcal{H} = - J \sum_{\langle i,j \rangle} (S_i^x S_j^x + S_i^y S_j^y),
\label{eq:Hamiltonian}
\end{equation}
where the summation runs over all nearest-neighbor pairs, $J = 1$ is the coupling constant, and $\mathbf{S}_i = (S_i^x, S_i^y, S_i^z)$ is a three-component spin vector at site $i$, subject to the constraint $(S_i^x)^2 + (S_i^y)^2 + (S_i^z)^2 = 1$. Since the Hamiltonian depends only on the planar components, this model is closely related to the planar-rotator model, in which spins are restricted to two components and satisfy $(S_i^x)^2 + (S_i^y)^2 = 1$. Due to these similarities and the fact that both exhibit a BKT transition, the XY model is often confused with the planar-rotator model in the literature~\cite{rotor1, rotor2}.

\section{Results}\label{sec:res}

To investigate the effects of the Padé approximation on each of the methods introduced, we first consider the two-dimensional Ising model with lattice size $L=24$, for which the exact density of states is known~\cite{exactising}. This provides a benchmark for directly evaluating the accuracy of the reconstructed Fisher zeros. Once validated, the Padé-based approach is applied to both the Ising and XY models to estimate the critical temperature and compared it with the original formulations. In addition, a runtime analysis is performed to quantify the computational gains associated with the reduction in the number of zeros.

For the XY model, the density of states was obtained using a Wang–Landau simulation. Details can be found in \cite{rochabkt}, and the raw data are available in \cite{dataxy}.

\subsection{Ising Model Benchmark}
\subsubsection{Fisher Zeros}\label{res:pade_fisher}

\begin{figure}[!ht]
    \centering
    \includegraphics{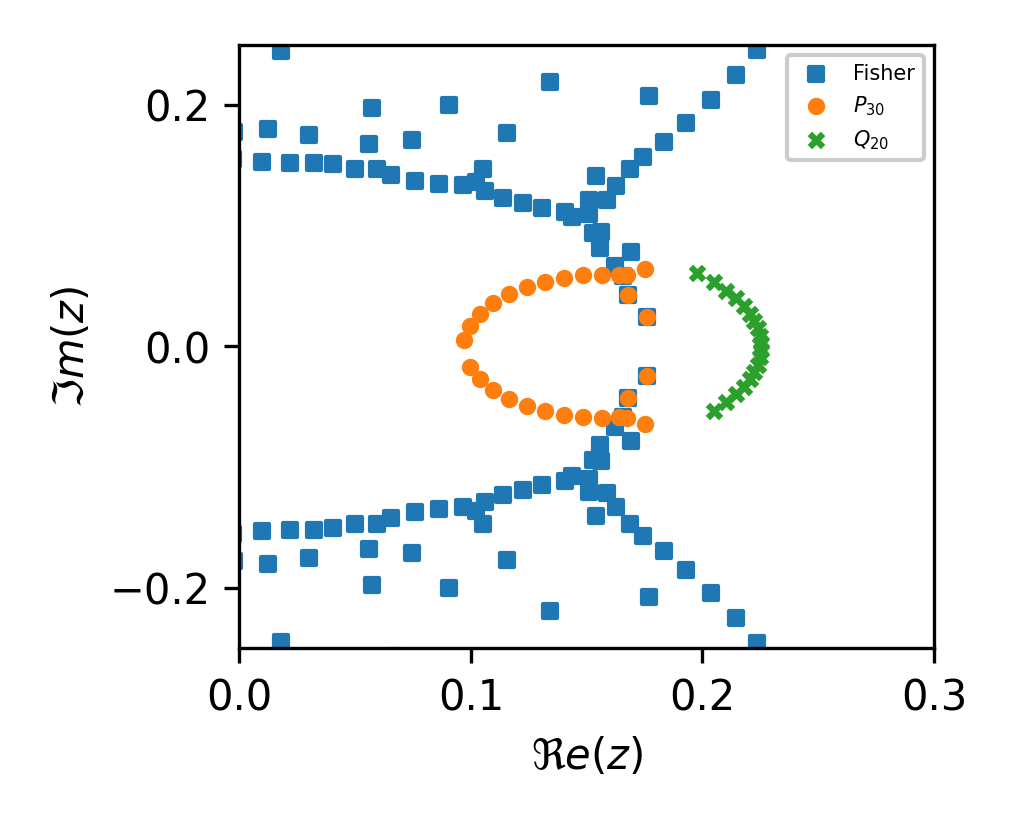}
    \caption{Fisher zeros for the Ising model with $L=24$. The exact Fisher zeros are shown alongside the Padé-Fisher approximation obtained via the shifted Padé method with $m=30$, $n=20$, and $r_0=0.17$. With only $30$ terms in the approximation, the dominant zero is correctly identified.}
    \label{fig:fisher_spade}
\end{figure}

To apply the Padé approximation to the Fisher zeros (Padé-Fisher), the density of states is used as input to generate the approximated zeros. The resulting distributions are shown in Fig.~\ref{fig:pade_fisher} for different values of $m$. An accurate identification of the dominant zero is already achieved for $m=150$, allowing the dominant zero to be determined using one quarter of the original polynomial order required for the full Fisher zeros map. Furthermore, with the shifted version of the Padé approximation, it was necessary only $m=30$ to find the dominant zero as can be seen in Fig.~\ref{fig:fisher_spade}.

\subsubsection{EPD Zeros}\label{res:pade_epd}

\begin{figure}[!ht]
    \centering
    \includegraphics{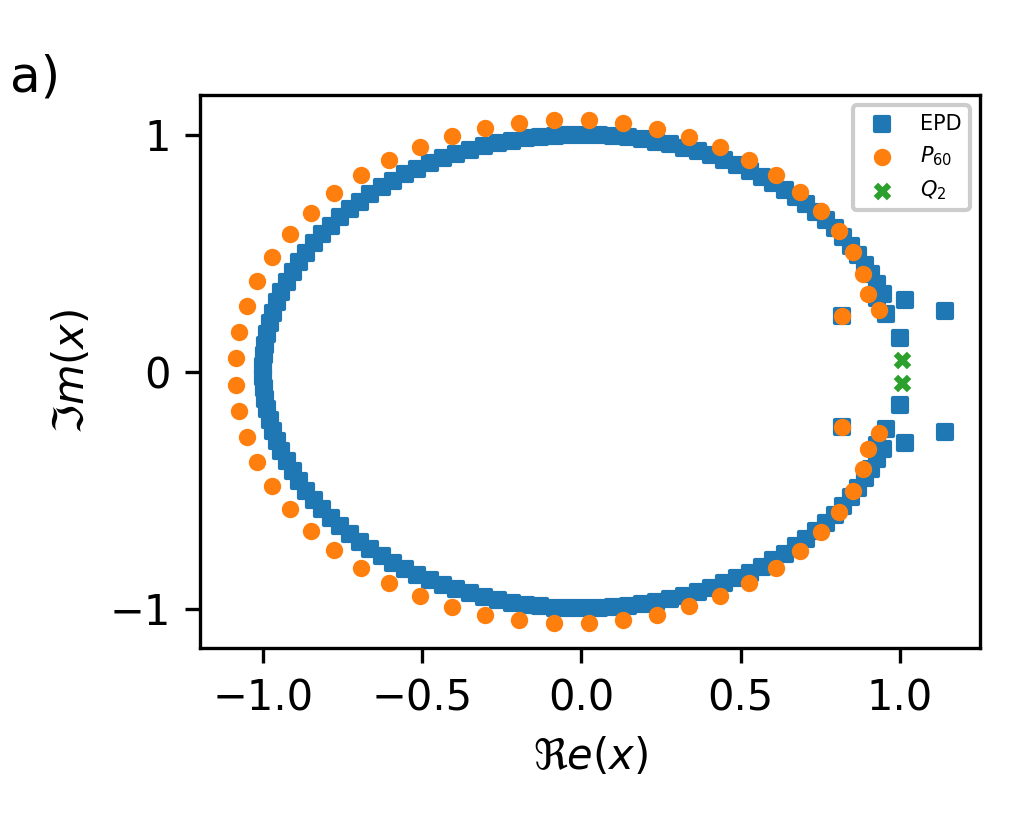}
    \includegraphics{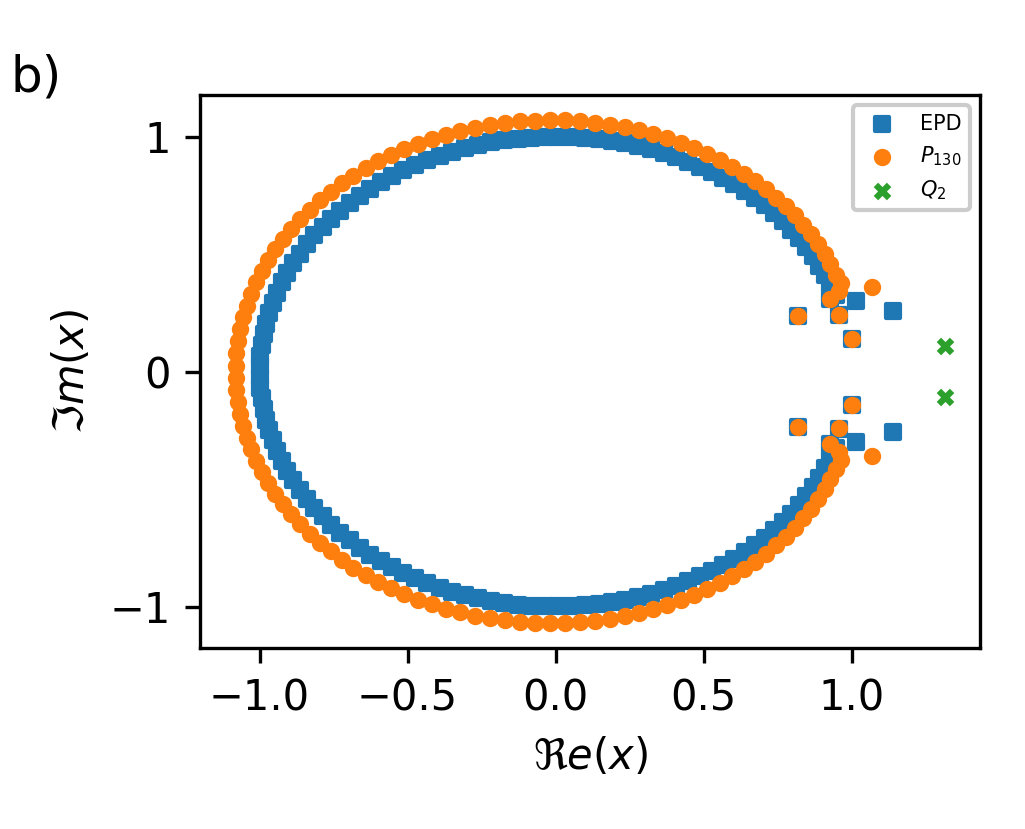}
    \caption{Distribution of EPD zeros for the Ising model with $h_t=10^{-4}$, $T=2.3$, and $L=24$, compared with those obtained from the Padé approximation. (a) For $m=60$ and $n=2$, a cluster of $Q_2$ zeros appears near the dominant zero, indicating insufficient resolution and the need for larger $m$. (b) For $m=130$, the dominant zero is correctly identified; however, the required polynomial order is comparable to that of the standard EPD method.}
    \label{fig:epd}
\end{figure}

Surprisingly, when the Padé approximation is applied to the EPD method (Padé-EPD), no significant advantage is observed. The value of $m$ required to obtain a reliable approximation of the dominant zero is nearly identical to the polynomial order used in the standard EPD method, making the Padé approximation effectively redundant. Fig. \ref{fig:epd} shows the distribution of zeros for $m = [60, 130]$ and $n = 2$, only for $m=130$ the dominant zero is correctly estimated, whereas the original EPD polynomial has degree $136$.

\subsubsection{MGF Zeros}\label{res:pade_mgf}

\begin{figure}
    \centering
    \includegraphics{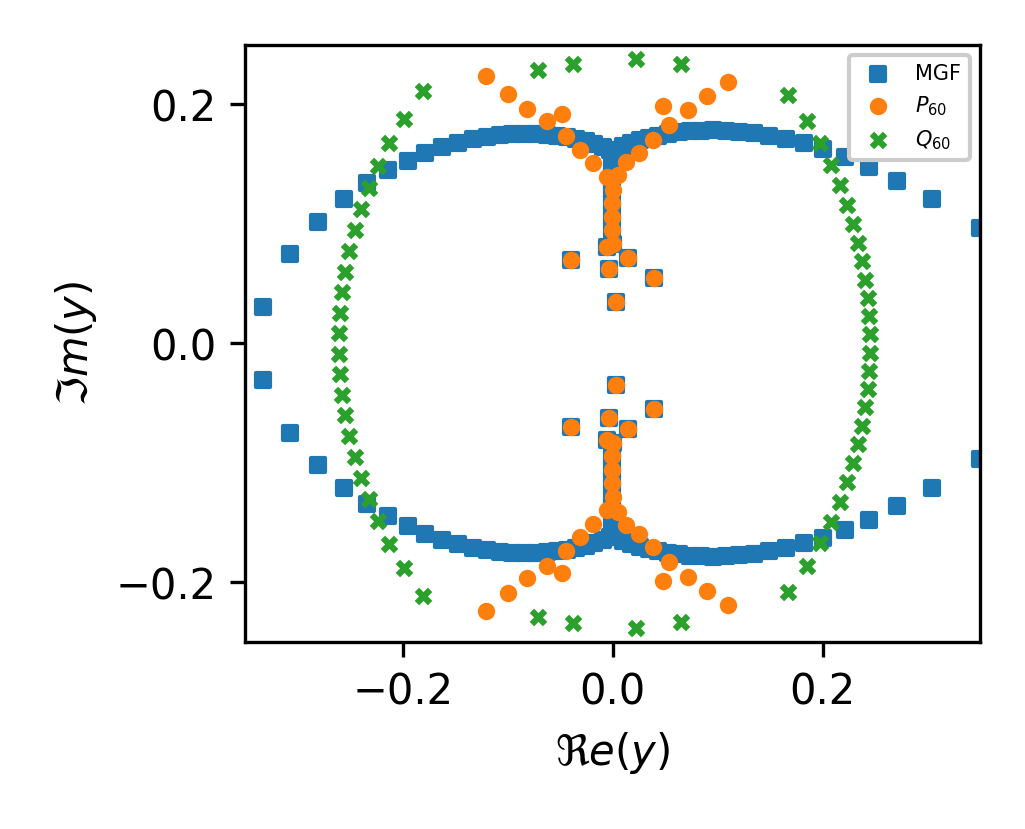}
    \caption{Distribution of MGF zeros for the Ising model with $L=24$ at $T=2.3$, using $k_{\text{max}}=120$, compared with those obtained from the Padé approximation. The Padé parameters are set to $m=60$ and $n=60$, corresponding to half the polynomial degree of the original MGF formulation.}
    \label{fig:mgf}
\end{figure}

The Padé approximation applied to the MGF formulation (Padé-MGF) proves to be particularly effective, as anticipated from its interpretation as a rational extension of a truncated Taylor expansion. As shown in Fig.~\ref{fig:mgf}, a Padé approximation constructed with approximately half the polynomial degree of the original MGF expansion is sufficient to accurately reproduce the dominant zero. Although smaller values of $m$ can still yield accurate results, we adopt $m$ equal to half the degree of the original MGF polynomial, as this value is sufficient to reproduce equivalent results.

\subsection{Method Limitations in the XY Model}\label{res:problemxy}

\begin{figure}
    \centering
    \includegraphics{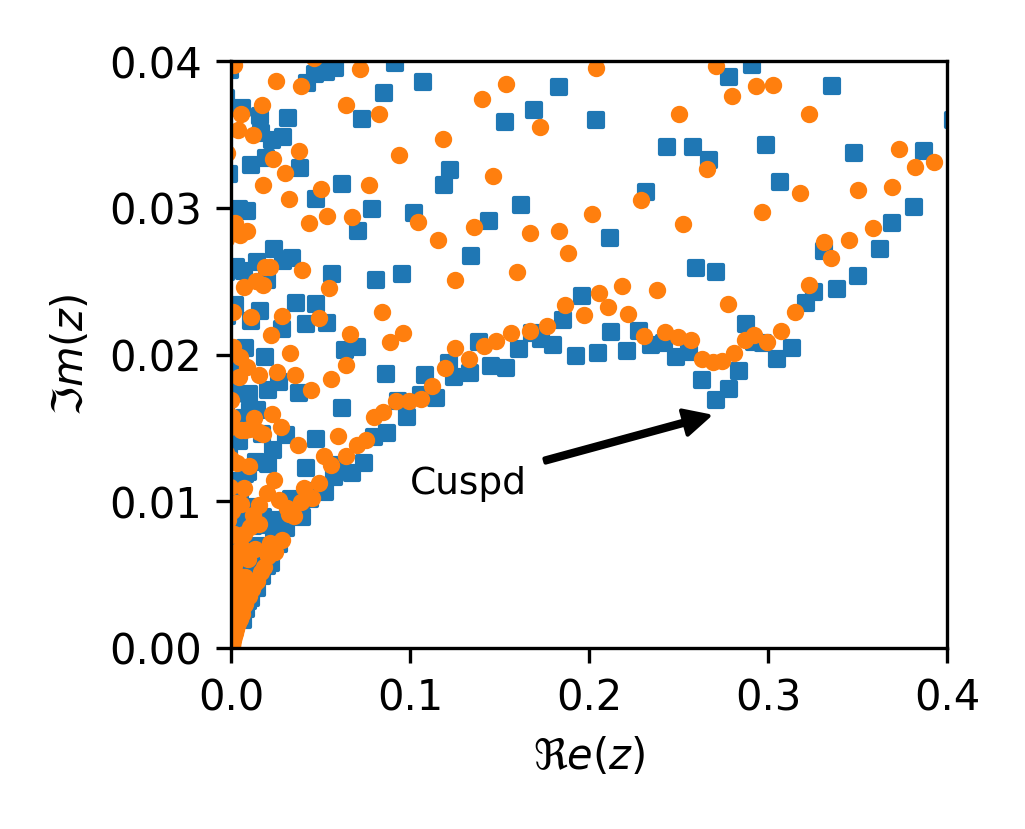}    
    \caption{Two samples of Fisher zeros for the XY model with lattice size $L=50$. The phase transition is identified from the cusp structure formed by the zeros in the complex plane. Notably, the cusp is not always sharply defined.}
    \label{fig:xy_fisher}
\end{figure}

When applying zeros-based methods to the XY model, fundamental differences emerge in comparison with the Ising case. Unlike the Ising model, the XY model does not exhibit a single well-defined dominant zero associated with the phase transition. Instead, the transition must be identified from the cusp formed by the internal boundary of zeros, see Fig. \ref{fig:xy_fisher}, combined with a finite-size scaling analysis as discussed in Ref.~\cite{rochabkt}.

\begin{figure}
    \centering
    \includegraphics{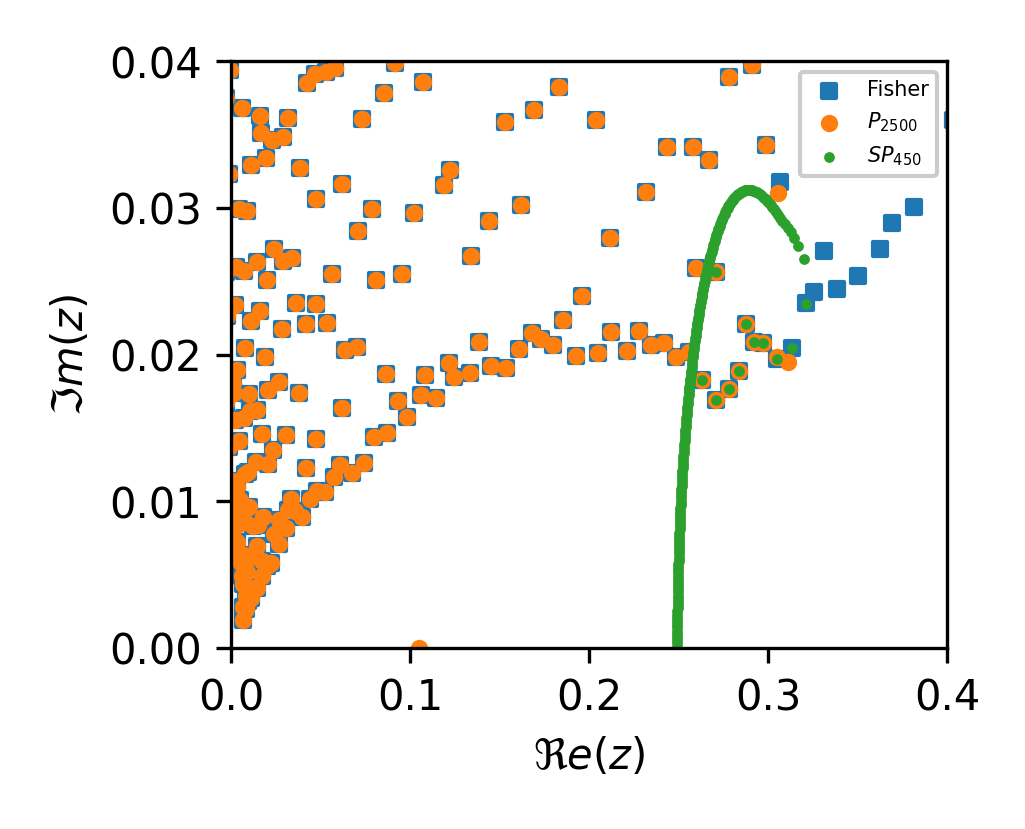}
    \caption{Zeros map for the Fisher, Padé, and shifted Padé (SP) approximations, with the shift taken at $r_0=0.27$. Note that the shifted Padé approximation captures only the region containing the cusp.}
    \label{fig:xy_cusp}
\end{figure}

\begin{figure}
    \centering
    \includegraphics{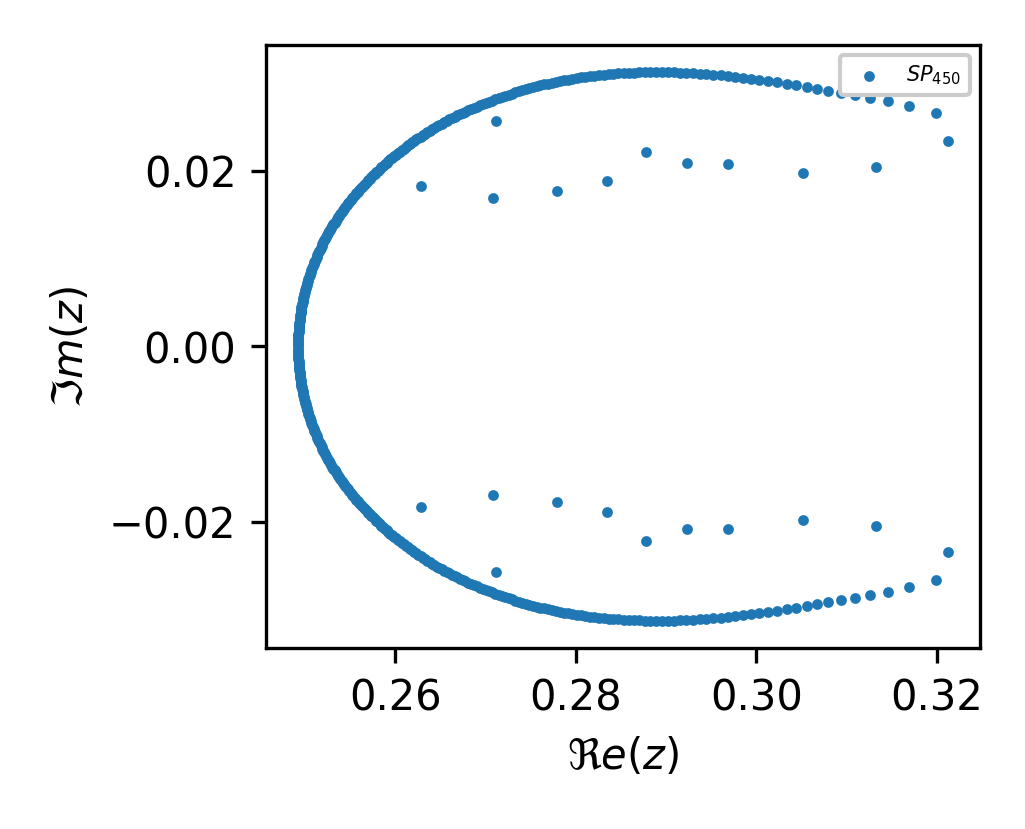}
    \caption{Zeros map for the shifted Padé approximation (SP), where the shift is taken at $r_0=0.27$. When only the shifted Padé zeros map is shown, it is not possible to precisely identify the location of the cusp.}
    \label{fig:xy_spade}
\end{figure}

The shifted Padé, MGF, and EPD approaches reconstruct only a local portion of the zeros map, as illustrated in Fig.~\ref{fig:xy_cusp}. This limited reconstruction is insufficient to unambiguously determine the cusp position due to the absence of global structural information, a limitation that becomes evident when the shifted Padé zeros are considered in isolation, as shown in Fig.~\ref{fig:xy_spade}. Furthermore, both the MGF and EPD methods exhibit convergence issues in the XY model, as discussed in ~\ref{sec:non_conv}, which further compromises their reliability. Therefore, the only method that can accurately locate the phase transition is the Padé–Fisher method, as it preserves the global structure while reducing the number of required zeros.

To enable a consistent comparison between the approaches, we applied the EPD, MGF, and shifted Padé methods to the XY model, using the Fisher zeros as a reference since these methods cannot independently locate the phase transition. The temperatures in the EPD and MGF analyses are set to the finite-size transition temperatures $T_{BKT}(L)$ obtained from the Fisher zeros, while the shift parameter $r_0$ is fixed at the cusp position for $L=50$. The resulting zeros are mapped onto the Fisher zeros plane (Fig.~\ref{fig:xy_cusp}) and analyzed following Ref.~\cite{rochabkt}, ensuring that all methods are evaluated near the cusp. 

\subsection{Critical Temperature and Number of Zeros}\label{res:models}

For finite lattice sizes, the partition function zeros yield pseudo-critical temperatures that depend on the system size. Consequently, a finite-size scaling analysis is necessary to reliably extrapolate the critical temperature in the thermodynamic limit. For the Ising model, the scaling follows
\begin{equation}
T_c(L) \sim T_c + L^{-1/\nu},
\end{equation}
with $\nu = 1$ in two dimensions. In contrast, the XY model exhibits BKT scaling \cite{rochabkt}, characterized by logarithmic corrections,
\begin{equation}
T_{\mathrm{BKT}}(L) \sim T_{\mathrm{BKT}} + \frac{1}{(\ln L)^{2}}.
\end{equation}

\begin{figure}
    \centering
    \includegraphics{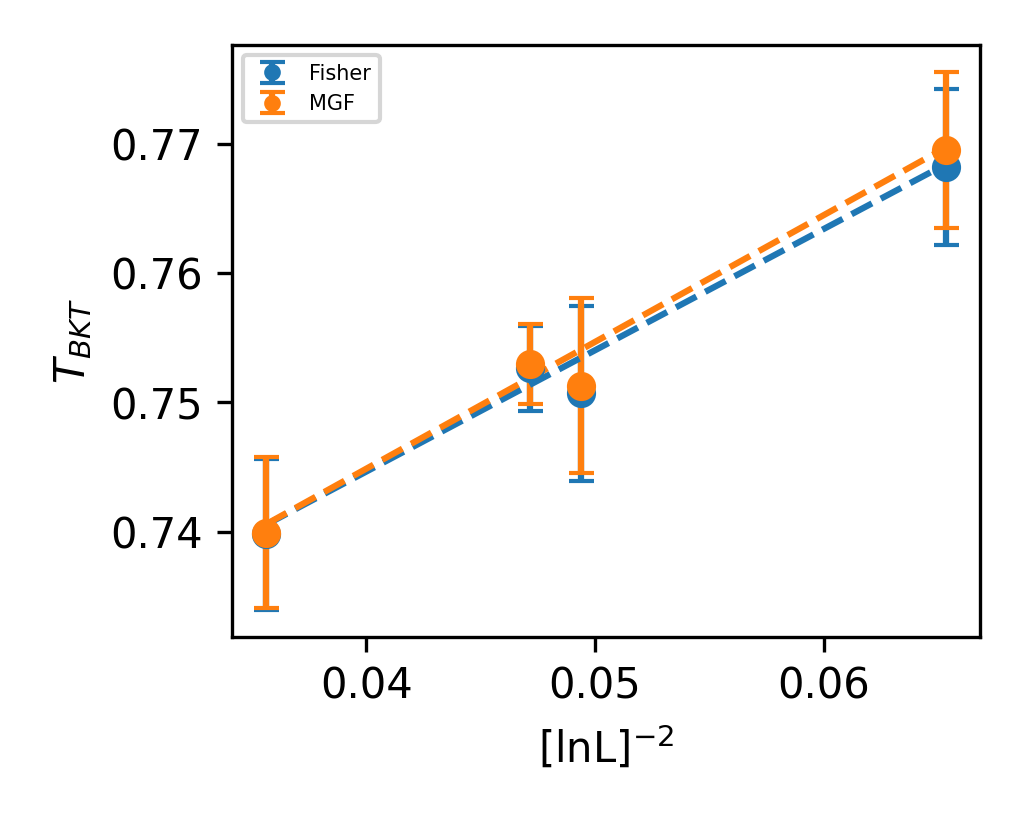}
    \caption{Finite-size scaling for the XY model as a function of $[\text{ln} L]^{-2}$ \cite{scalingxylog}, using system sizes $L=[50, 90, 100, 200]$. For clarity, only the linear regressions obtained from the Fisher and MGF zeros are shown. The regression lines corresponding to the other methods coincide exactly with one of these two and therefore lie directly on top of them. Table~\ref{tab:nTc} summarizes the numerical values of $T_c$ for each method.}
    \label{fig:regression_xy}
\end{figure}

The finite-size scaling analysis for the XY model is shown in Fig.~\ref{fig:regression_xy}, based on multiple DOS datasets obtained via Wang–Landau simulations~\cite{dataxy}. For each system size $L$, the pseudo-critical temperature $T_c(L)$ is computed for each sample, from which the mean and standard deviation are calculated.

The resulting critical temperatures from the different methods are summarized in Table~\ref{tab:nTc}. The shifted Padé approach is not compatible with the convergence schemes of the EPD and MGF methods and is therefore not considered, while the Padé-EPD results are omitted due to the absence of improvement over the standard formulation. All estimates are consistent with exact results for the Ising model and with previously reported values for the XY model.

\begin{table}[!ht]
    \centering
    \setlength{\tabcolsep}{10pt}
    \renewcommand{\arraystretch}{1.2}
    \begin{tabular}{lcc}
        \hline
        \multicolumn{3}{c}{\textbf{Critical Temperature}} \\
        \hline
        \textbf{Method} & Ising & XY  \\
        Fisher         &      2.27098(8)     &     0.707(7)            \\
        Padé--Fisher   &      2.27098(8)     &     0.707(7)            \\
        SPadé--Fisher   &      2.27098(8)     &     0.707(7)            \\
        EPD   &      2.27098(8)     &     0.707(7)            \\
        MGF            &      2.26799(5)  &   0.705(7)             \\
        Padé--MGF      &     2.26799(5)     &     0.705(7)            \\
    \end{tabular}
    \caption{Critical temperature for the Ising model and XY model for each method. The label SPadé is used to denote the shifted Padé approximation. For the Ising model, the exact critical temperature is $T_c=2.269185$. For the XY model, literature estimates of the transition temperature are approximately $0.700$ \cite{PhysRevB.54.12302,rochabkt,epd}.}
    \label{tab:nTc}
\end{table}

The reduction in polynomial degree and, consequently, in the number of zeros is illustrated in Tables \ref{tab:nzeros} for the Ising and XY models. For the Ising model with $L=150$, the Padé approximation reduces the number of Fisher zeros from $22{,}500$ to $5{,}000$, while in the MGF formulation the number of zeros is reduced by half. As a direct consequence, the computational cost of finding the polynomial roots using \texttt{MPSolve} decreases dramatically: for instance, in the $L=150$ Ising lattice, the complete Fisher zeros computation required about $34$ minutes \footnote{Time measured on a Dell Inspiron system (Intel i7-7700HQ, 16 GB RAM) using a Python implementation with the \texttt{mpmath} library for arbitrary precision arithmetic.}, whereas the Padé method required only $80$ seconds. The shifted Padé method further reduces the number of zeros to $150$, with a total runtime of approximately $3$ minutes.

A similar trend is observed in the XY model with $L=200$, where the full Fisher zeros map required $3.46$ hours of computation, while its Padé counterpart was completed in approximately $1$ hour, and the shifted Padé version further decreased it to only $21$ minutes. In contrast to the Ising case, the shifted Padé approach is more efficient here, which can be attributed to the significantly larger number of zeros in the XY model ($ 68{,}000$) compared to the Ising model ($22{,}500$).

These results demonstrate that the Padé approximation preserves the physical content of the original methods while significantly improving computational performance. The zeros obtained from the Padé-based formulations coincide with those from the full polynomial, confirming that the approximation introduces no measurable deviation in the estimation of the critical temperature.

\begin{table}[!ht]
    \centering
    \setlength{\tabcolsep}{10pt}
    \renewcommand{\arraystretch}{1.2}
    \begin{tabular}{lcccc}
        \hline
        \multicolumn{5}{c}{\textbf{Ising Model}} \\
        \hline
        \textbf{Method} & $L{=}24$ & $L{=}32$ & $L{=}64$ & $L{=}150$ \\
        Fisher         &      576     &     1024      &      4096     &    22500       \\
        Padé--Fisher   &      150     &     1000      &    2500       &    5000       \\
        SPadé--Fisher   &      30     &     30      &    50       &    150       \\
        EPD   &      136     &      193     &    429       &     1687      \\
        MGF            &      120  &   120        &  120         &  120         \\
        Padé--MGF      &     60      &     60      &     60      &    60       \\
    \end{tabular}
    \centering
    \setlength{\tabcolsep}{10pt}
    \renewcommand{\arraystretch}{1.2}
    \begin{tabular}{lcccc}
        \hline
        \multicolumn{5}{c}{\textbf{XY Model}} \\
        \hline
        \textbf{Method} & $L{=}50$ & $L{=}90$ & $L{=}100$ & $L{=}200$ \\
        Fisher         &      4751     &      14580     &      18001     &    68000       \\
        Padé--Fisher   &        2500   &     7700      &    9500       &    36000       \\  
        SPadé--Fisher   &        350   &     900      &    1000       &    1500       \\  
        EPD   &      559     &     948      &   1089        &    2092       \\
        MGF            &      320  &   320        &  320         &  320         \\
        Padé--MGF      &     160      &     160      &     160      &    160       \\
    \end{tabular}
    \caption{Number of zeros required to determine the critical temperature for the Ising model and XY model. The number of zeros increases with system size $L$ for the EPD and Fisher related methods, whereas for the MGF related methods it remains constant.}
    \label{tab:nzeros}
\end{table}

\section{\label{sec:conc} Closing Remarks}

The results demonstrate that the Padé approximation constitutes an efficient and accurate method for analyzing phase transitions via partition function zeros. When applied to the Fisher and MGF formulations, it retains the physical content of the original methods while substantially reducing the polynomial degree required to identify the dominant zero, leading to a significant decrease in computational cost without loss of precision. On the other hand, its application to the EPD zeros does not offer any improvement, as the polynomial order required to find the dominant zero is similar in both methods.

The method was validated using the two-dimensional Ising and anisotropic Heisenberg (XY) models, yielding critical temperatures in agreement with the literature values while substantially reducing the number of required zeros. For the largest Ising lattice considered, the Padé-Fisher approach reduced the number of zeros from $25{,}000$ to $5{,}000$, decreasing the computation time from $34$ minutes to $80$ seconds. In the XY model, the number of Fisher zeros was reduced from $68{,}000$ to $36{,}000$, with the computation time decreasing from $3.46$ hours to approximately one hour. The shifted Padé variant further decreases the computational cost by restricting the analysis to a localized region of the zeros map, requiring only $150$ zeros for the Ising model and approximately $1{,}000$ for the XY model, with runtimes of about $3$ and $21$ minutes, respectively.

Although all considered methods provide accurate estimates of the critical temperature in systems with a well-defined dominant zero, the EPD, MGF, and shifted Padé approaches are not suitable for a direct analysis of the XY model. The EPD and MGF formulations suffer from intrinsic convergence issues, preventing a reliable identification of the transition, while all three methods reconstruct only a localized portion of the zeros map, thereby failing to capture the global cusp structure characteristic of the XY transition. In contrast, the Padé approximation applied to the Fisher zeros preserves the global features of the zeros distribution, providing a robust and reliable method for analyzing the XY model.

\section*{Data Availability}

The raw data used in the XY model can be found in Zenodo \cite{dataxy}.

\section*{Acknowledgements}

This study was financed by the São Paulo Research Foundation (FAPESP), Brasil. Process Number $2023/15458$-$4$.

\appendix
\section{Non-Convergent Behavior of Zeros in the XY Model}\label{sec:non_conv}

\begin{figure}
    \centering
    \includegraphics{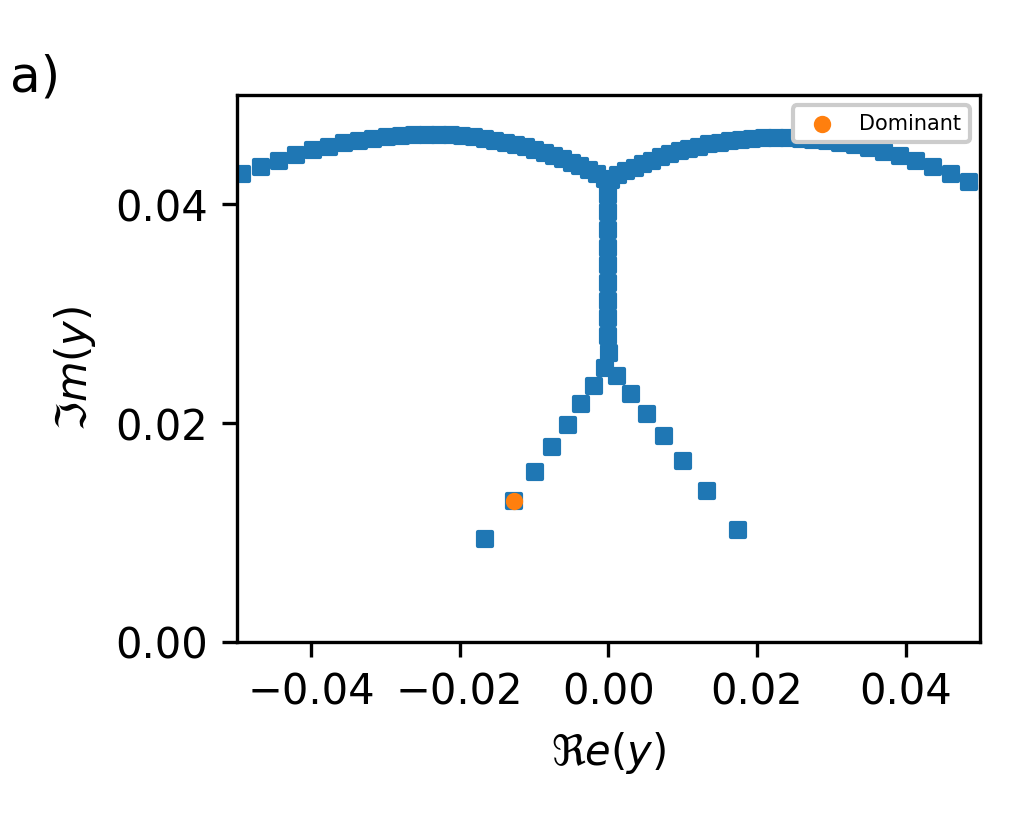}
    \includegraphics{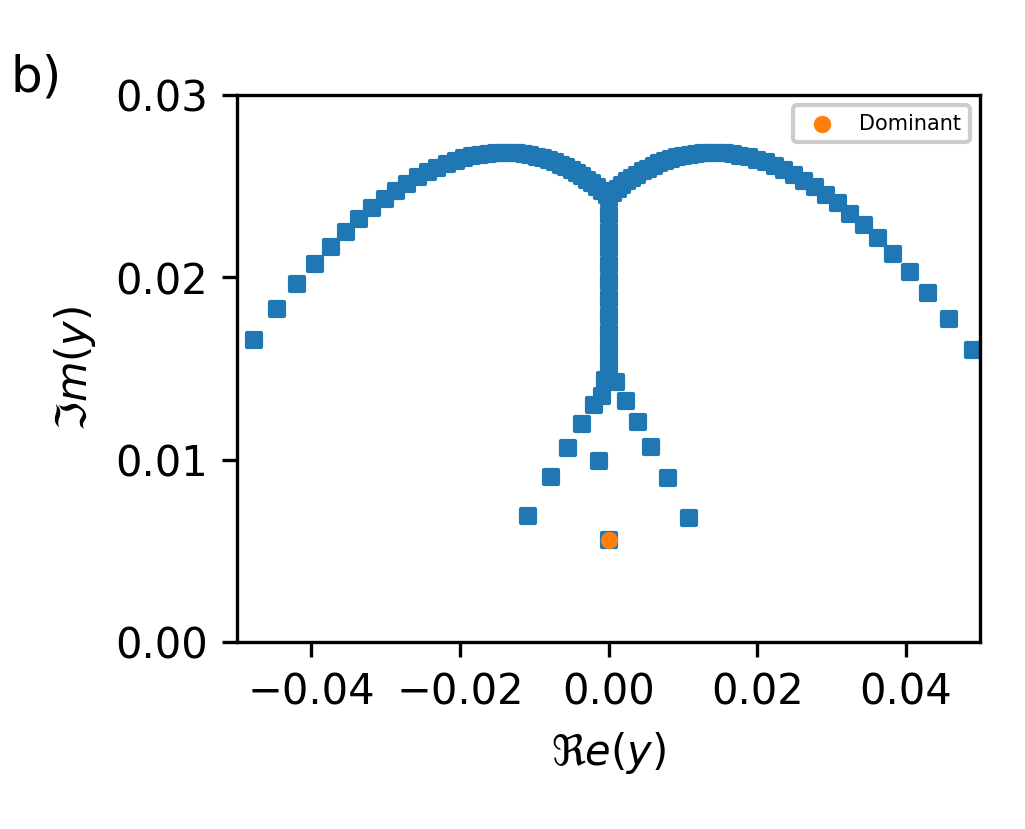}
    \caption{MGF zeros for the Ising model with lattice size $L = 150$ at (a) $T = 3.0$ and (b) $T_c = 2.265$. At the critical temperature, the dominant zero is clearly identifiable.}
    \label{fig:ising_mgf}
\end{figure}

\begin{figure}
    \centering
    \includegraphics{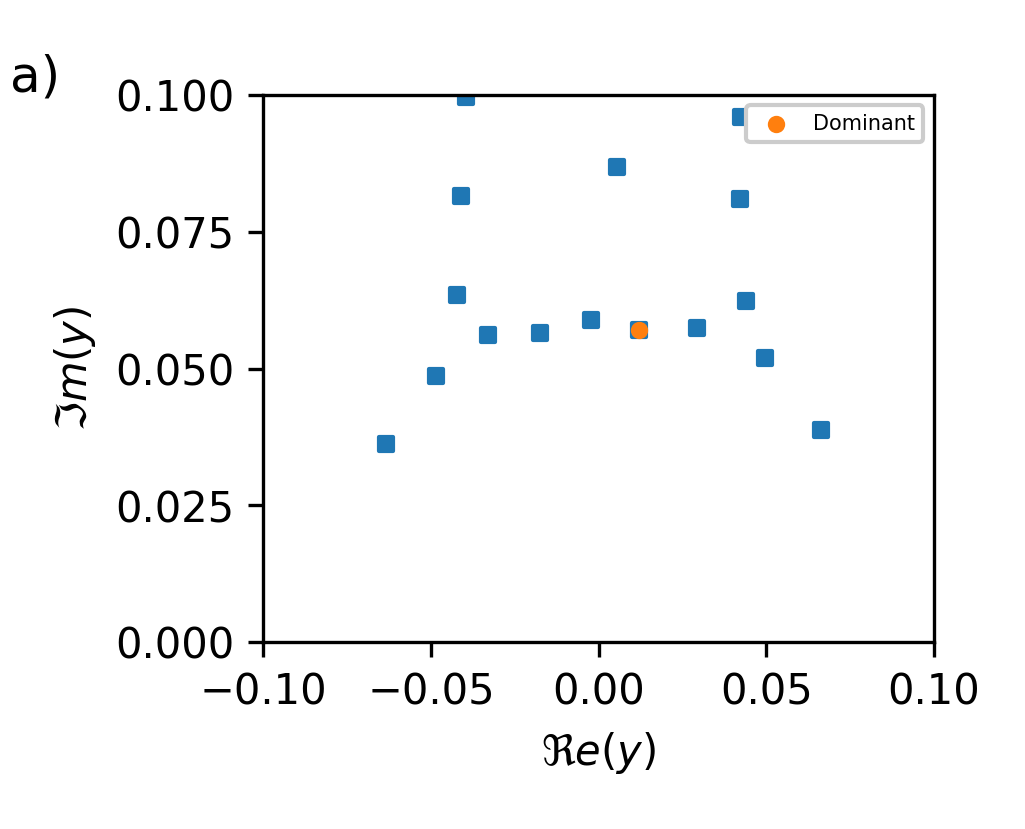}
    \includegraphics{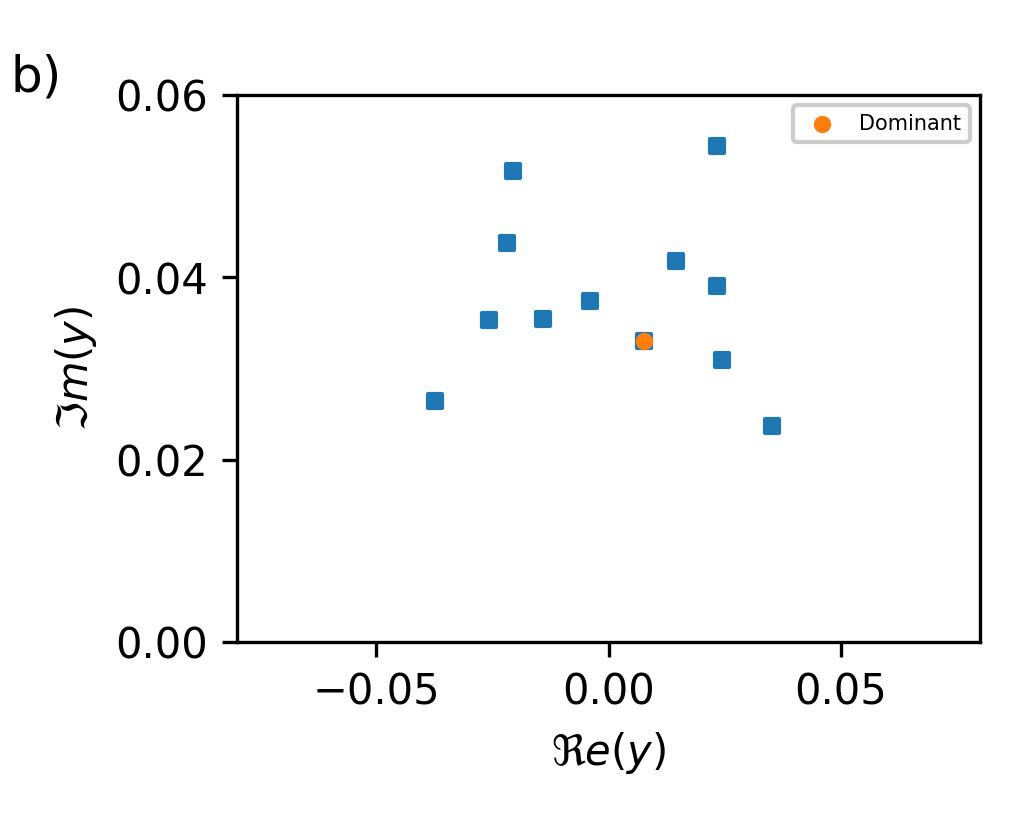}
    \caption{MGF zeros for the XY model with lattice size $L = 100$ at (a) $T = 2.0$ and (b) $T_{\text{BKT}} = 0.7587$. In contrast to the Ising case, the zeros do not exhibit a distinct dominant zero; instead, several zeros appear near $(0,0)$, which may be incorrectly identified as the dominant zero by the convergence algorithm.}
    \label{fig:xy_mgf}
\end{figure}

In contrast to the Ising case, see Fig.~\ref{fig:ising_mgf}, where a well-defined dominant zero is observed, the XY model exhibits an extended line of zeros in the region where a single dominant zero would normally be expected, even for temperatures far from the critical point, see Fig.~\ref{fig:xy_mgf}. As a consequence of the continuous line of zeros for the XY model, the convergence algorithms discussed in this work tend to approach temperatures close to any initial value $\beta_o$ leading to a wrong estimation of $T_{BKT}(L)$. It should be noted that convergence is observed for small lattice sizes, yielding estimates consistent with the expected critical temperature. However, this behavior does not persist for larger systems.

\bibliography{mybibfile}

@article{mgf,
  title = {Moment-generating function zeros in the study of phase transitions},
  author = {Rodrigues, R. G. M. and Costa, B. V. and M\'ol, L. A. S.},
  journal = {Phys. Rev. E},
  volume = {104},
  issue = {6},
  pages = {064103},
  numpages = {7},
  year = {2021},
  month = {Dec},
  publisher = {American Physical Society},
  doi = {10.1103/PhysRevE.104.064103},
  url = {https://link.aps.org/doi/10.1103/PhysRevE.104.064103}
}

@article{epd,
title = "Energy probability distribution zeros: A route to study phase transitions",
journal = "Computer Physics Communications",
volume = "216",
pages = "77 - 83",
year = "2017",
issn = "0010-4655",
doi = "https://doi.org/10.1016/j.cpc.2017.03.003",
url = "http://www.sciencedirect.com/science/article/pii/S0010465517300796",
author = "B.V. Costa and L.A.S. Mól and J.C.S. Rocha",
}

@incollection{fisherzeros,
  author      = {M. E. Fisher},
  title       = {The Nature of Critical Points},
  editor      = {Brittin, W.E.},
  booktitle   = {Lectures in Theoretical Physics},
  publisher   = {University of Colorado Press, Boulder},
  year        = {1965},
  pages       = {1-159},
  volume={7C},
  chapter     = {1},
}

@article{cumulants1,
  title = {Lee-Yang theory, high cumulants, and large-deviation statistics of the magnetization in the Ising model},
  author = {Deger, Aydin and Brange, Fredrik and Flindt, Christian},
  journal = {Phys. Rev. B},
  volume = {102},
  issue = {17},
  pages = {174418},
  numpages = {12},
  year = {2020},
  month = {Nov},
  publisher = {American Physical Society},
  doi = {10.1103/PhysRevB.102.174418},
  url = {https://link.aps.org/doi/10.1103/PhysRevB.102.174418}
}

@article{cumulants2,
  title = {Lee-Yang theory of the Curie-Weiss model and its rare fluctuations},
  author = {Deger, Aydin and Flindt, Christian},
  journal = {Phys. Rev. Res.},
  volume = {2},
  issue = {3},
  pages = {033009},
  numpages = {10},
  year = {2020},
  month = {Jul},
  publisher = {American Physical Society},
  doi = {10.1103/PhysRevResearch.2.033009},
  url = {https://link.aps.org/doi/10.1103/PhysRevResearch.2.033009}
}

@article{cumulants,
  title = {Determination of universal critical exponents using Lee-Yang theory},
  author = {Deger, Aydin and Flindt, Christian},
  journal = {Phys. Rev. Res.},
  volume = {1},
  issue = {2},
  pages = {023004},
  numpages = {7},
  year = {2019},
  month = {Sep},
  publisher = {American Physical Society},
  doi = {10.1103/PhysRevResearch.1.023004},
  url = {https://link.aps.org/doi/10.1103/PhysRevResearch.1.023004}
}

@article{yanglee1,
  title = {Statistical Theory of Equations of State and Phase Transitions. I. Theory of Condensation},
  author = {Yang, C. N. and Lee, T. D.},
  journal = {Phys. Rev.},
  volume = {87},
  issue = {3},
  pages = {404--409},
  numpages = {0},
  year = {1952},
  month = {Aug},
  publisher = {American Physical Society},
  doi = {10.1103/PhysRev.87.404},
  url = {https://link.aps.org/doi/10.1103/PhysRev.87.404}
}

@article{yanglee2,
  title = {Statistical Theory of Equations of State and Phase Transitions. II. Lattice Gas and Ising Model},
  author = {Lee, T. D. and Yang, C. N.},
  journal = {Phys. Rev.},
  volume = {87},
  issue = {3},
  pages = {410--419},
  numpages = {0},
  year = {1952},
  month = {Aug},
  publisher = {American Physical Society},
  doi = {10.1103/PhysRev.87.410},
  url = {https://link.aps.org/doi/10.1103/PhysRev.87.410}
}

@article{rochabkt,
title = "Using zeros of the canonical partition function map to detect signatures of a Berezinskii–Kosterlitz–Thouless transition",
journal = "Computer Physics Communications",
volume = "209",
pages = "88 - 91",
year = "2016",
issn = "0010-4655",
doi = "https://doi.org/10.1016/j.cpc.2016.08.016",
url = "http://www.sciencedirect.com/science/article/pii/S0010465516302466",
author = "J.C.S. Rocha and L.A.S. Mól and B.V. Costa",
}

@article{tensorxylj,
author = {Hong ,Seongpyo and Kim ,Dong-Hee},
title = {Tensor Network Calculation of the Logarithmic Correction Exponent in the XY Model},
journal = {Journal of the Physical Society of Japan},
volume = {91},
number = {8},
pages = {084003},
year = {2022},
doi = {10.7566/JPSJ.91.084003},

URL = { 
    
        https://doi.org/10.7566/JPSJ.91.084003
    
    

},
eprint = { 
    
        https://doi.org/10.7566/JPSJ.91.084003
    
    

}
}

@article{Moueddene_2024,
doi = {10.1088/1742-5468/ad1d60},
url = {https://doi.org/10.1088/1742-5468/ad1d60},
year = {2024},
month = {feb},
publisher = {IOP Publishing},
volume = {2024},
number = {2},
pages = {023206},
author = {Moueddene, Leïla and G Fytas, Nikolaos and Holovatch, Yurij and Kenna, Ralph and Berche, Bertrand},
title = {Critical and tricritical singularities from small-scale Monte Carlo simulations: the Blume–Capel model in two dimensions},
journal = {Journal of Statistical Mechanics: Theory and Experiment}
}

@article{Moueddene_2025,
doi = {10.1088/1742-5468/ae09a4},
url = {https://doi.org/10.1088/1742-5468/ae09a4},
year = {2025},
month = {oct},
publisher = {IOP Publishing},
volume = {2025},
number = {10},
pages = {104001},
author = {Moueddene, Leïla and G Fytas, Nikolaos and Berche, Bertrand},
title = {Phase transition properties via partition function zeros: the Blume–Capel ferromagnet revisited},
journal = {Journal of Statistical Mechanics: Theory and Experiment}
}

@article{PhysRevB.54.12302,
  title = {Critical dynamics in the two-dimensional classical XY model: A spin-dynamics study},
  author = {Evertz, Hans Gerd and Landau, D. P.},
  journal = {Phys. Rev. B},
  volume = {54},
  issue = {17},
  pages = {12302--12317},
  numpages = {0},
  year = {1996},
  month = {Nov},
  publisher = {American Physical Society},
  doi = {10.1103/PhysRevB.54.12302},
  url = {https://link.aps.org/doi/10.1103/PhysRevB.54.12302}
}

@article{Costa2019,
author = {Costa, B. V. and M{\'{o}}l, L. A.S. and Rocha, J. C.S.},
doi = {10.1007/s13538-019-00636-x},
issn = {16784448},
journal = {Brazilian Journal of Physics},
number = {2},
pages = {271--276},
title = {{A New Algorithm to Study the Critical Behavior of Topological Phase Transitions}},
volume = {49},
url = {https://link.springer.com/article/10.1007/s13538-019-00636-x},
year = {2019}
}

@article{Lima2019,
author = {Lima, A. B. and M{\'{o}}l, L. A. S. and Costa, B. V.},
doi = {10.1007/s10955-019-02271-x},
issn = {0022-4715},
journal = {Journal of Statistical Physics},
month = {jun},
number = {5},
pages = {960--971},
title = {{The Fully Frustrated XY Model Revisited: A New Universality Class}},
url = {http://link.springer.com/10.1007/s10955-019-02271-x},
volume = {175},
year = {2019}
}

@article{fisherzerosbosegas,
author = {van Dijk, Wytse and Lobo, Calvin and MacDonald, Allison and Bhaduri, Rajat K.},
title = {Fisher zeros of a unitary Bose gas},
journal = {Canadian Journal of Physics},
volume = {93},
number = {8},
pages = {830-835},
year = {2015},
doi = {10.1139/cjp-2014-0585},
URL = {https://doi.org/10.1139/cjp-2014-0585
},
eprint = {https://doi.org/10.1139/cjp-2014-0585},
}

@article{graphzerosising,
	doi = {10.1088/1751-8113/49/13/135001},
	url = {https://doi.org/10.1088/1751-8113/49/13/135001},
	year = 2016,
	month = {feb},
	publisher = {{IOP} Publishing},
	volume = {49},
	number = {13},
	pages = {135001},
	author = {M Krasnytska and B Berche and Yu Holovatch and R Kenna},
	title = {Partition function zeros for the Ising model on complete graphs and on annealed scale-free networks},
	journal = {Journal of Physics A: Mathematical and Theoretical},
}

@article{zerosmeasurequantsimul,
  title = {Measuring complex-partition-function zeros of Ising models in quantum simulators},
  author = {Krishnan, Abijith and Schmitt, Markus and Moessner, Roderich and Heyl, Markus},
  journal = {Phys. Rev. A},
  volume = {100},
  issue = {2},
  pages = {022125},
  numpages = {8},
  year = {2019},
  month = {Aug},
  publisher = {American Physical Society},
  doi = {10.1103/PhysRevA.100.022125},
  url = {https://link.aps.org/doi/10.1103/PhysRevA.100.022125}
}

@article{expzerosyl,
  title = {Experimental Observation of Lee-Yang Zeros},
  author = {Peng, Xinhua and Zhou, Hui and Wei, Bo-Bo and Cui, Jiangyu and Du, Jiangfeng and Liu, Ren-Bao},
  journal = {Phys. Rev. Lett.},
  volume = {114},
  issue = {1},
  pages = {010601},
  numpages = {5},
  year = {2015},
  month = {Jan},
  publisher = {American Physical Society},
  doi = {10.1103/PhysRevLett.114.010601},
  url = {https://link.aps.org/doi/10.1103/PhysRevLett.114.010601}
}

@article{benareview,
   title={STATISTICAL MECHANICS OF EQUILIBRIUM AND NONEQUILIBRIUM PHASE TRANSITIONS: THE YANG–LEE FORMALISM},
   volume={19},
   ISSN={1793-6578},
   url={http://dx.doi.org/10.1142/S0217979205032759},
   DOI={10.1142/s0217979205032759},
   number={29},
   journal={International Journal of Modern Physics B},
   publisher={World Scientific Pub Co Pte Lt},
   author={BENA, IOANA and DROZ, MICHEL and LIPOWSKI, ADAM},
   year={2005},
   month={Nov},
   pages={4269–4329}
}

@article{dynamicquant,
  title = {Dynamical Quantum Phase Transitions in the Transverse-Field Ising Model},
  author = {Heyl, M. and Polkovnikov, A. and Kehrein, S.},
  journal = {Phys. Rev. Lett.},
  volume = {110},
  issue = {13},
  pages = {135704},
  numpages = {5},
  year = {2013},
  month = {Mar},
  publisher = {American Physical Society},
  doi = {10.1103/PhysRevLett.110.135704},
  url = {https://link.aps.org/doi/10.1103/PhysRevLett.110.135704}
}

@article{rochapol,
title = {Leading Fisher Partition Function Zeros as Indicators of Structural Transitions in Macromolecules},
journal = {Physics Procedia},
volume = {57},
pages = {94-98},
year = {2014},
note = {Proceedings of the 27th Workshop on Computer Simulation Studies in Condensed Matter Physics (CSP2014)},
issn = {1875-3892},
doi = {https://doi.org/10.1016/j.phpro.2014.08.139},
url = {https://www.sciencedirect.com/science/article/pii/S1875389214002843},
author = {Julio C.S. Rocha and Stefan Schnabel and David P. Landau and Michael Bachmann},
}

@book{40yearskt,
  title={40 Years Of Berezinskii--kosterlitz--thouless Theory},
  author={Jose, J.V. and Thouless, D.},
  isbn={9789814417655},
  url={https://books.google.com.br/books?id=OfC6CgAAQBAJ},
  year={2013},
  publisher={World Scientific Publishing Company}
}

@article{kt1,
doi = {10.1088/0022-3719/6/7/010},
url = {https://doi.org/10.1088/0022-3719/6/7/010},
year = {1973},
month = {apr},
publisher = {},
volume = {6},
number = {7},
pages = {1181},
author = {J M Kosterlitz and D J Thouless},
title = {Ordering, metastability and phase transitions in two-dimensional systems},
journal = {Journal of Physics C: Solid State Physics}
}

@article{kt2,
doi = {10.1088/0022-3719/5/11/002},
url = {https://doi.org/10.1088/0022-3719/5/11/002},
year = {1972},
month = {jun},
publisher = {},
volume = {5},
number = {11},
pages = {L124},
author = {J M Kosterlitz and D J Thouless},
title = {Long range order and metastability in two dimensional solids and superfluids. (Application of dislocation theory)},
journal = {Journal of Physics C: Solid State Physics}
}

@article{nolongrangeorderkt,
  title = {Absence of Ferromagnetism or Antiferromagnetism in One- or Two-Dimensional Isotropic Heisenberg Models},
  author = {Mermin, N. D. and Wagner, H.},
  journal = {Phys. Rev. Lett.},
  volume = {17},
  issue = {22},
  pages = {1133--1136},
  numpages = {0},
  year = {1966},
  month = {Nov},
  publisher = {American Physical Society},
  doi = {10.1103/PhysRevLett.17.1133},
  url = {https://link.aps.org/doi/10.1103/PhysRevLett.17.1133}
}

@article{rotor1,
doi = {10.1088/0305-4470/38/26/003},
url = {https://doi.org/10.1088/0305-4470/38/26/003},
year = {2005},
month = {jun},
publisher = {},
volume = {38},
number = {26},
pages = {5869},
author = {Hasenbusch, Martin},
title = {The two-dimensional XY model at the transition temperature: a high-precision Monte Carlo study},
journal = {Journal of Physics A: Mathematical and General}
}

@article{rotor2,
  title = {Monte Carlo analysis of the two-dimensional XY model. II. Comparison with the Kosterlitz renormalization-group equations},
  author = {Olsson, Peter},
  journal = {Phys. Rev. B},
  volume = {52},
  issue = {6},
  pages = {4526--4535},
  numpages = {0},
  year = {1995},
  month = {Aug},
  publisher = {American Physical Society},
  doi = {10.1103/PhysRevB.52.4526},
  url = {https://link.aps.org/doi/10.1103/PhysRevB.52.4526}
}

@article{qcdcumulutant,
  title = {Exploring Lee-Yang and Fisher zeros in the 2D Ising model through multipoint Pad\'e approximants},
  author = {Singh, Simran and Cipressi, Massimo and Di Renzo, Francesco},
  journal = {Phys. Rev. D},
  volume = {109},
  issue = {7},
  pages = {074505},
  numpages = {13},
  year = {2024},
  month = {Apr},
  publisher = {American Physical Society},
  doi = {10.1103/PhysRevD.109.074505},
  url = {https://link.aps.org/doi/10.1103/PhysRevD.109.074505}
}

@article{scalingxylog,
title = {Logarithmic corrections to scaling in the two dimensional XY-model},
journal = {Physics Letters B},
volume = {351},
number = {1},
pages = {273-278},
year = {1995},
issn = {0370-2693},
doi = {https://doi.org/10.1016/0370-2693(95)00316-D},
url = {https://www.sciencedirect.com/science/article/pii/037026939500316D},
author = {R. Kenna and A.C. Irving}
}

@book{numrec,
author = {Press, William H. and Teukolsky, Saul A. and Vetterling, William T. and Flannery, Brian P.},
title = {Numerical Recipes 3rd Edition: The Art of Scientific Computing},
year = {2007},
isbn = {0521880688},
publisher = {Cambridge University Press},
address = {USA},
edition = {3}
}

@article{exactising,
  title = {Exact Distribution of Energies in the Two-Dimensional Ising Model},
  author = {Beale, Paul D.},
  journal = {Phys. Rev. Lett.},
  volume = {76},
  issue = {1},
  pages = {78--81},
  numpages = {0},
  year = {1996},
  month = {Jan},
  publisher = {American Physical Society},
  doi = {10.1103/PhysRevLett.76.78},
  url = {https://link.aps.org/doi/10.1103/PhysRevLett.76.78}
}

@dataset{dataxy,
  author       = {Siqueira Rocha, Julio Cesar and
                  Mól, Lucas and
                  Costa, Bismarck V},
  title        = {Entropy per Energy for the XY Model},
  month        = jun,
  year         = 2025,
  publisher    = {Zenodo},
  doi          = {10.5281/zenodo.15614663},
  url          = {https://doi.org/10.5281/zenodo.15614663},
}

@software{BaileyMPFUN,
  author = {David H. Bailey},
  title  = {MPFUN2020: A thread-safe arbitrary precision package
with special functions},
  year   = {2020},
  url    = {https://www.davidhbailey.com/dhbsoftware/}
}

\end{document}